\documentclass{iopart}
\usepackage{graphicx}
\usepackage{eqnarray}
\usepackage{iopams}
\usepackage{cite}
\usepackage{color}

\begin{document}
\title{Observations of Pauli Paramagnetic Effects on the Flux Line Lattice in CeCoIn$_{5}$}
\author{J S White$^{1}$~\footnote{Present Address: Laboratory for Neutron Scattering, Paul Scherrer Institut, CH-5232 Villigen PSI, Switzerland.}, P Das$^{2}$, M R Eskildsen$^{2}$,\\ L DeBeer-Schmitt$^{2}$~\footnote{Present Address: Oak Ridge National Laboratory, MS 6393 Oak Ridge, TN 37831-6393, USA.}, E M Forgan$^{1}$, A D Bianchi$^{3}$,\\ M Kenzelmann$^{4}$, M Zolliker$^{4}$, S Gerber$^{5}$, J L Gavilano$^{5}$,\\ J Mesot$^{5,6}$, R Movshovich$^{7}$, E D Bauer$^{7}$, J L Sarrao$^{7}$ and\\ C Petrovic$^{8}$}
\address{$^{1}$ School of Physics and Astronomy, University of Birmingham, Edgbaston, Birmingham, B15 2TT, UK.}
\address{$^{2}$ Department of Physics, University of Notre Dame, Notre Dame, IN 46556, USA.}
\address{$^{3}$ D\'{e}partement de Physique and Regroupement Qu\'{e}b\'{e}cois sur les Mat\'{e}riaux de Pointe, Universit\'{e} de Montr\'{e}al, Montr\'{e}al, QC, H3C 3J7, Canada.}
\address{$^{4}$ Laboratory for Developments and Methods, Paul Scherrer Institute, CH-5232 Villigen, Switzerland.}
\address{$^{5}$ Laboratory for Neutron Scattering, ETH Z\"{u}rich and Paul Scherrer Institut, CH-5232 Villigen, Switzerland.}
\address{$^{6}$ Institut de Physique de la Mati\`{e}re Complexe, EPFL, CH-1015 Lausanne, Switzerland.}
\address{$^{7}$ Los Alamos National Laboratory, Los Alamos, NM 87545, USA.}
\address{$^{8}$ Brookhaven National Laboratory, Upton, NY 11973, USA.}
\ead{jonathan.white@psi.ch}

\begin{abstract}
\\From small-angle neutron scattering studies of the flux line lattice (FLL) in CeCoIn$_{5}$, with magnetic field applied parallel to the crystal \textbf{c}-axis, we obtain the field- and temperature-dependence of the FLL form factor, which is a measure of the spatial variation of the field in the mixed state. We extend our earlier work [A.D. Bianchi~\emph{et al.} 2008 Science \textbf{319} 177] to temperatures up to 1250~mK. Over the entire temperature range, paramagnetism in the flux line cores results in an increase of the form factor with field. Near $H_{c2}$ the form factor decreases again, and our results indicate that this fall-off extends outside the proposed FFLO region. Instead, we attribute the decrease to a paramagnetic suppression of Cooper pairing. At higher temperatures, a gradual crossover towards more conventional mixed state behavior is observed.
\end{abstract}

\section{Introduction}
The heavy-fermion superconductor CeCoIn$_{5}$ continues to excite great interest, because it shows strong Pauli paramagnetic effects~\cite{Bia02,Tay02} and also the close proximity of superconductivity to a quantum critical point~\cite{Bia03a,Pag03}. It has a superconducting transition temperature, $T_{c}\sim 2.3$~\textrm{K} in zero field~\cite{Pet01}, with a $d_{x^{2}-y^{2}}$ order-parameter~\cite{Iza01,Vor06}. At low temperature, the transition to the normal state is first-order~\cite{Bia02,Tay02}, showing that the superconductivity is suppressed by coupling of the field to the anti-parallel spins of the singlet Cooper pair (the Pauli effect) rather than the more common coupling to the orbital motion of Cooper pairs in the mixed state (the orbital effect)~\cite{Clo62,Mak66}. The dominance of the Pauli effect, combined with the quasi-2D structure and super-clean crystal properties ($\ell \sim 1000 \xi$~\cite{Orm02}), means that the stringent requirements for the stabilisation of the inhomogeneous Fulde-Ferrell-Larkin-Ovchinnikov (FFLO) state are satisfied~\cite{Ful64,Lar64,Mat07}. Numerous studies report experimental signatures that are compatible with the formation of FFLO~\cite{Bia03b,Rad03,Kak05,Mit08,Kum06}, though an unambiguous microscopic observation remains elusive. Recently, both nuclear magnetic resonance (NMR)~\cite{You07} and neutron diffraction studies~\cite{Ken08} have provided microscopic evidence for the stabilisation of field-induced long range antiferromagnetic order for fields applied within the basal plane. This magnetically ordered phase (termed `$Q$-phase' in~\cite{Ken08}) exists only within the superconducting mixed state, and occupies the same high field, low temperature region of the superconducting $(H,T)$ phase diagram as proposed for FFLO. It remains unclear whether this ordered phase replaces, or coexists with, a non-standard FFLO state.

%Recently, an antiferromagnetically ordered `$Q$-phase' has been observed for magnetic fields parallel to [110], and only within the superconducting mixed state~\cite{Ken08}. The $Q$-phase exists in the same high field, low temperature region of the mixed state $(H,T)$ phase diagram as proposed for FFLO. It remains unclear whether the $Q$-phase replaces, or coexists with, a non-standard FFLO state.

Small-angle neutron scattering (SANS) studies of the superconducting mixed state of CeCoIn$_{5}$ have determined the flux line lattice (FLL) structure and orientation for the major part of the superconducting phase diagram with magnetic field applied parallel to the crystal \textbf{c}-axis $\left(\textbf{H}\parallel\textbf{c}\right)$~\cite{Esk03,DeB06,Bia08,Ohi08}. In addition, SANS can be used to determine the FLL form factor (FF) which is a measure of the field contrast in the mixed state, i.e. the difference between the local magnetic induction in the flux line cores and at the field-minima between them~\cite{Esk03,DeB06,Bia08,Ohi08}. Most notably, an anomalous field-driven \emph{enhancement} of the FF was observed at 50~mK upon approaching $H_{c2}(T=0)=4.95$~T~\cite{Bia08}. We proposed that the increasing FF arises from paramagnetic moments induced in the flux line cores where the antiparallel alignment of electron spins in Cooper pairs is suppressed, a phenomenon that is consistent with the predictions of recent numerical calculations~\cite{Ich07b}. Just before entering the normal state, an abrupt fall of the FF was observed, which we speculated was associated with the formation of the FFLO state~\cite{Bia08}. However, the numerical study~\cite{Ich07b} reproduced a similar reduction of the FF at high field, apparently unrelated to the onset of the FFLO state. Hence, the physical origin of the decreasing FF just below $H_{c2}$ remains unclear. To address this question, we have used SANS to investigate the field-dependence of the FLL FF in CeCoIn$_{5}$ for temperatures up to 1250~mK. For all temperatures in this range, the FF initially rises with increasing field, reaching a maximum before it starts to fall prior to entering the normal state. Our results show that while we are unable to rule out a contribution to the field-dependence of the FF due to a FFLO-type state, the dominant contribution to the high field fall in the FF at low temperature is due to a paramagnetic suppression of the Cooper pairing.

\section{Experimental}
Our experiments were performed on the SANS-I instrument at the Paul Scherrer Institut. The samples were mosaics of co-aligned CeCoIn$_{5}$ plate-like single crystals (with small dimension along the \textbf{c}-axis) grown from an excess indium flux~\cite{Pet01}. Most results reported here were obtained using a sample of total volume $0.2 \times 11 \times 14$~mm$^3$ and of total mass 250~mg. Thin samples were necessary due to the strong neutron absorption of indium. The sample was mounted in a dilution refrigerator with base temperature $\sim50$~mK which was inserted into a horizontal field cryomagnet, and oriented so that the \textbf{c}-axis was $\lesssim 2^{\circ}$ from the field direction. Neutrons of wavelengths $\lambda_{n}=5-6$~\AA~were selected with a wavelength spread $\Delta\lambda_{n}/\lambda_{n} = 10\%$, and were incident approximately parallel to the field; the diffracted neutrons were collected by a position-sensitive multidetector. At all fields the FLL was prepared by field-cooling through $T_{c}$. Diffraction measurements were performed as the sample and cryomagnet were rotated together to carry the FLL diffraction spots through the Bragg condition. Background subtraction was performed using measurements taken either above $T_{c}$ or after zero-field cooling, with no noticeable dependence on which method was used.

\section{Results and Discussion}
The local field in the mixed state may be expressed as a sum over its spatial Fourier components with indices $(h,k)$, and scattering vectors $\textrm{\textbf{q}}_{hk}$ belonging to the two-dimensional FLL reciprocal lattice. The form factor (FF) at wavevector $\textrm{\textbf{q}}_{hk}$ is the magnitude of the Fourier component $F(\textrm{\textbf{q}}_{hk})$. The value of the FF is obtained from the integrated intensity of a FLL Bragg reflection as the FLL is rotated through the diffraction condition. This integrated intensity, $I_{hk}$, is related to the modulus squared of the FF, $|F(\textrm{\textbf{q}}_{hk})|^{2}$, by~\cite{Chr77}
\begin{equation}\label{1}
    I_{hk} = 2\pi V\phi \left(\frac{\gamma}{4}\right)^{2} \frac{\lambda_{n}^{2}}{\Phi_{0}^2 \; q_{hk}}|F(\textrm{\textbf{q}}_{hk})|^{2}.
\end{equation}
Here, $V$ is the illuminated sample volume, $\phi$ is the incident neutron flux density, $\lambda_{n}$ is the neutron wavelength, $\gamma$ is the magnetic moment of the neutron in nuclear magnetons, and $\Phi_{0} = h/2e$ is the flux quantum. General predictions from Ginzburg-Landau (GL) theory, whether by numerical methods or algebraic approximations, are for a monotonic decrease of the FLL FF with increasing field~\cite{Cle75,Yao97,Hao91,Bra95}. Although strictly valid only close to $T_{c}$, these models have been widely used and give a good description of results from many superconductors at lower temperatures~\cite{Yar97,Kea00,For02,Cub03,Gil04,Cub07,Den09}. Quasiclassical theory, which is valid well away from $T_{c}$, gives qualitatively similar results for both $s$- and $d$-wave orbitally-limited superconductors~\cite{Ich99}. We will see that the conventional picture provided by all of these theories lies in stark contrast to the results that we now report.

%Models based on Ginzburg-Landau (GL) theory give the field variation of the FF in terms of the characteristic length-scales for the superconductor, the penetration depth and the coherence length~\cite{Cle75,Yao97,For02}. The general prediction made using GL theory is for a monotonic decrease of the FF with increasing field, in stark contrast to the results that we now report.

In figure~1 we show the field-dependence of the FLL FF for first-order reflections, at various temperatures between 50 and 1250~mK. At all temperatures, the FF initially rises with increasing field before reaching a maximum, and then begins to fall again on approaching $H_{c2}$. At temperatures of 500~mK and below, the FF remains finite all the way up to $H_{c2}$, where the FLL signal disappears abruptly upon entering the normal state. This is consistent with the first-order nature of the superconducting transition seen in thermodynamic measurements at low temperatures~\cite{Bia02,Tay02}, and predicted theoretically for strong Pauli-limiting~\cite{Mak66}. Above 500~mK, the field-dependence of the FF becomes increasingly conventional, smoothly approaching zero at $H_{c2}$, as expected for a second-order transition to the normal state. However, even at these higher temperatures, Pauli-paramagnetic effects remain important, leading to a maximum in the FF at intermediate fields.

%This anisotropy is further found to be independent of field and temperature, and

Figure~2 shows the regions within the superconducting phase diagram occupied by the various FLL structures~\cite{Bia08}. For the hexagonal structures, two of the six Bragg spots in each of the two domains lie along $\langle110\rangle$ directions. In the high-field region, we find that the value of $|F(\textbf{\textrm{q}}_{10})|^{2}$ for these spots is $\sim~20\%$ larger than for any of the other four spots. This FF anisotropy is consistent with that expected to arise due to the slight distortion of the hexagonal structure away from the isotropic hexagon~\cite{Bia08}. As a consequence of this, the magnitude of \textbf{q}$_{10}$ is \emph{always} slightly less for the two $\langle110\rangle$ aligned Bragg spots than for the other four spots. To accommodate this variation in intensity, the FF values shown in figure~1 are an average over all six spots. Whenever the FLL structure changes, both the magnitudes and directions of $\textbf{q}_{hk}$ alter; however, figure~1 shows that any changes in the value of $|F(\textbf{\textrm{q}}_{10})|^{2}$ at the FLL structure phase boundaries are no bigger than the errors. Finally in figure~1, we see that at low field there is a common value of FF, independent of temperature, showing the differences between the data at different temperatures occur at higher fields. We may interpret this behavior as showing that the effects of Pauli limiting are small at low fields, and that for $T\lesssim T_{c}/2$, orbital-limiting effects are fairly temperature-independent. However, at high fields, Pauli-limiting is dominant and the proximity of $H_{c2}(T)$ causes major changes in the behaviour of the FF.

In figure~3 we show the results of investigating the detailed temperature-dependence of the FF in the high field hexagonal FLL structure phase. For the fields of 4.60~T, 4.85~T and 4.90~T, we have recorded the temperature-dependence of the diffracted intensity of a first-order Bragg reflection. At the two higher fields, these temperature scans pass through $H_{c2}$ almost parallel to the upper critical field phase boundary; nonetheless, there is a sharp fall in intensity as the first-order boundary is crossed, with perhaps a slight smearing arising from tiny differences between crystallites in the mosaic. We estimate that the contribution due to demagnetisation effects to the width of the discontinuity at $T_{c2}$ is unimportant. At the lower field, a slower variation with temperature of the normalised intensity is seen, reflecting the second-order transition at $H_{c2}$. However, as shown by the absolute data in the inset, the FF is at its maximum at the lower field, deeper within the mixed state, and falls as either field or temperature is increased towards $H_{c2}$.

Previous discussion of the field-induced amplification of the FF emphasised the inability of conventional theory to explain our results at low temperatures~\cite{Bia08}. Here, we consider the FF behavior in the context of an extension~\cite{Ich07b} to the quasi-classical Eilenberger theory~\cite{Eil68}. In this work, Pauli paramagnetic effects are incorporated by adding a Zeeman energy term $\mu B(\textrm{\textbf{r}})$ to the Eilenberger equations~\cite{Ich07b}, where the parameter $\mu$ represents the relative strength of the paramagnetism~\cite{Not09}. This theory was first used to successfully describe the similar, but less extreme, field dependence of the FF in TmNi$_2$B$_2$C~\cite{DeB07}. At low temperatures, and for large values of $\mu$, the model predicts not only an initial increase of the FF with increasing field, but also a decrease close to $H_{c2}$~\cite{Ich07b}. To obtain a reasonable qualitative agreement between the theory and the observed FF field-dependence in CeCoIn$_{5}$ requires $\mu\sim2.6$. However, experimentally determined material parameters for CeCoIn$_{5}$ imply a $\mu\simeq1.7$ for $\textrm{\textbf{H}} \parallel \textrm{\textbf{c}}$. Hence, for a \emph{quantitative} agreement with our data, some refinement of the theory is required. For instance, it might be necessary to incorporate a more three-dimensional character of the Fermi surface of CeCoIn$_{5}$~\cite{Set01}, or account for the non-Fermi liquid behavior which is most prominent near $H_{c2}$. We further mention that heat capacity measurements~\cite{Ike01} reveal a field-dependence to the low energy density of states (DOS), which further needs to be accounted for~\cite{Ich09}. The inclusion of a field-dependent DOS would be consistent with the observation by thermal conductivity studies of two-band superconductivity in CeCoIn$_{5}$~\cite{Sey08}, which imply that some of the supercarriers become depaired at very low fields. These depaired carriers would not only be available to contribute to the paramagnetic response at high fields, but their absence at low fields might also explain why the form factor at very low fields is not given by the expected expression (see~\cite{Bia08}) using the \emph{zero-field} value of the London penetration depth, $\lambda_{L}$~\cite{Orm02,Ozc03}. An independent bulk measurement of $\lambda_{L}$ at finite fields would test this suggestion.

It was previously suggested that the FF maximum at 50~mK could mark the onset of the FFLO state~\cite{Bia08}. Fig.~2 shows the temperature dependence of the field at which a maximum in the FF is observed, superposed onto the FLL structure phase diagram. It is clear that the FF maximum does not follow the FFLO phase boundary proposed experimentally~\cite{Bia03b,Kum06} or predicted theoretically~\cite{Gru66,Ada03}. Furthermore the FF maximum does not appear to be correlated with the high-field rhombic to hexagonal FLL symmetry transition. To understand the observed FF behavior close to $H_{c2}$, we therefore return to the the theory of Ichioka and Machida~\cite{Ich07b}, which reproduces the observed drop of the FF without including any effects of an FFLO state. Beginning at low fields, they predict an increase in the magnitude and spatial extent of the paramagnetic moment in the flux line cores, arising from the spin-split quasiparticle states located there. The spatially varying paramagnetic moment adds to the orbital contribution to the field variation in the mixed state to give an increase in the field contrast and hence in $F(\textbf{q}_{10})$. The size of the flux line cores, as measured by the extent of the region with a suppressed order parameter, also increases, showing the effects of paramagnetic depairing in these regions. However, at high fields where the width of the expanded flux line cores becomes comparable with the flux line spacing, the suppression of the order parameter allows the paramagnetic moment to spread further through the entire flux lattice unit cell. This leads to an additional paramagnetic moment of the \emph{whole crystal} which is observed in magnetisation measurements~\cite{Tay02}, but to a reduction in the \emph{field contrast}, as measured by SANS, and hence to a reduction in $F(\textbf{q}_{10})$. We emphasise that these effects are a consequence of paramagnetic effects dominating pairing in the near-core regions where the order parameter is suppressed, and that these effects spread out from there. Thus the decrease of the FF close to $H_{c2}$ is quite different in origin from the behavior seen in conventional materials.

Next we discuss the abrupt fall of the FF seen at $H_{c2}$ and low temperature. The finite value of the FF at $H_{c2}$ can be directly related to the jump in the magnetisation at the upper critical field~\cite{Tay02}. First, we note that the spatial field modulation $B(r)= \sum F(\textbf{q}_{hk}) \, \textrm{e}^{i \textbf{q}_{hk} \cdot \textbf{r}}$ where the sum is over all FLL reciprocal lattice indices $h$ and $k$, and $F(\textbf{q}=0)$ is equal to the average induction $\langle B \rangle$. Close to $H_{c2}$ it is a good approximation to include in the sum only the dominant Fourier component $F(\textbf{q}_{\{10\}})$, which is the FF measured in our SANS experiments. In the case of CeCoIn$_{5}$ it is important to repeat that the field modulation contains contributions from both the orbital screening currents and the polarisation of the unpaired electron spins. However, just below $H_{c2}$, the spin magnetisation is by far the dominant contribution to the field modulation, such that $B(\textbf{r}) = \mu_0 [M(\textbf{r}) + H_{\textrm{appl.}}]$~\cite{Ich09}. Assuming that the magnetisation in the ``normal" flux line cores is equal to $M_{\textrm{normal}}$ just above $H_{c2}$, the average magnetisation in the mixed state is easily found to be $\langle \mu_0 M \rangle = M_{\textrm{normal}} - 6 |F(\textbf{q}_{10})|$. This is illustrated in Fig. 4. Our measured value of the FF of 0.4~mT just below $H_{c2}$ at 50~mK would thus correspond to a jump in magnetisation of $\Delta (\mu_0 M)$ = 2.4~mT as the field is increased through the upper critical field. This is in very satisfactory agreement with the results of bulk magnetisation measurements, which give a jump at $H_{c2}$ of 3~mT~\cite{Tay02}.

%Their result may be understood in terms of an expansion of the flux line cores, and an increased core overlap, which leads to a reduction of the magnetic field contrast close to $H_{c2}$, despite the fact that the paramagnetic effects increase monotonically with field. Such a flux line core expansion is predicted both for larger values of $\mu$, and with increasing field for constant $\mu$~\cite{Ich09}. The size of the core is reflected both in the order parameter suppression and in the spatial extent of the paramagnetic magnetization, which is the dominant contribution to the magnetization at high field.

%One expected consequence of a flux line core expansion is that the FF for higher-order Bragg reflections from the FLL would be even more strongly suppressed than the first-order ones. We have observed $\textbf{q}_{11}$ reflections in the square and rhombic FLL phases, but in the hexagonal phase they were too weak to be confirmed under similar counting conditions. In the hexagonal phase, the higher-order intensities will be determined partly by the relatively larger value of $\textbf{q}_{11}$, and partly by the core size. From simple modeling of our data, we conclude that our observations are \emph{consistent} with core expansion at high fields. However, to give a \emph{measure} of the core structure a high fields, actual observation of higher order reflections close to $H_{c2}$ would be of great interest.

We now consider the effects of a flux line core paramagnetism on the sequence of FLL structures shown in Fig.~2, for fields $H > 0.5 H_{c2}$ and low temperature. We note that this sequence of transitions from square to rhombic and then to high-field distorted hexagonal, is the exact reverse of what is observed for $H < 0.5 H_{c2}$. In the central square phase, the FLL nearest neighbor directions are aligned with the nodes of the $d_{x^{2}-y^{2}}$ gap; this is consistent with the low-field sequence of transitions being driven by flux line core anisotropy, which reflects that of the order parameter. However, on the approach to $H_{c2}$, paramagnetic depairing causes the flux line cores to begin to expand and also become more isotropic~\cite{Ich09}. We expect that this suppresses the anisotropy which stabilised the square phase, and that core expansion is the driving mechanism for the high-field sequence of FLL structures. This may also explain the similar sequence of FLL transitions observed in TmNi$_2$B$_2$C~\cite{Esk98}, where Pauli paramagnetic effects have also been demonstrated~\cite{DeB07}. At higher temperatures, although the high field effects of paramagnetic depairing on both the FLL FF and the FLL structure become suppressed, figure~2 shows that the re-entrant square to rhombic transition persists to temperatures above those for which $H_{c2}$ is first-order. This is a further demonstration that paramagnetic effects remain important at higher temperatures where $H_{c2}(T)$ is becoming orbitally-limited. These deductions are consistent with those of another recent study based on a GL type theory, that also reproduces the observed sequence of high-field transitions and ascribes them primarily to the effects of paramagnetic de-pairing and Fermi surface anisotropy~\cite{Hia08}. However, we point out that GL theory is not expected to be numerically accurate over this region of the $(H,T)$ phase diagram.

\section{Conclusion}
In conclusion, we have studied the detailed field- and temperature-dependence of the diffracted neutron intensity from the flux line lattice in CeCoIn$_{5}$. At high fields and low temperatures, we most clearly observe the effects of Pauli paramagnetism in the flux line cores, which results in an increase of the FLL form factor that extends well into the temperature region where $H_{c2}$ becomes second order. Close to $H_{c2}$ the flux line lattice form factor decreases, which we attribute to the effects of a paramagnetic suppression of the Cooper pairing. We suggest that further consequences of this paramagnetic de-pairing are manifested in the observed high field sequence of FLL structure transitions.

%We attribute the field dependence of the form factor and the high field flux line lattice structure phase transitions to an expansion of the flux line cores, which occurs as paramagnetic de-pairing becomes increasingly significant at high fields.

\section{Acknowledgements}
We acknowledge valuable discussions with M.~Ichioka and K.~Machida. Experiments were performed at the Swiss spallation neutron source SINQ, Paul Scherrer Institut,  Villigen, Switzerland. We acknowledge support from the EPSRC of the UK, the U. S. NSF through grant DMR-0804887, the Alfred P. Sloan Foundation, NSERC (Canada), FQRNT (Qu\'{e}bec), the Canada Research Chair Foundation, the Swiss National Centre of Competence in Research program ``Materials with Novel Electronic Properties'', and from the European Commission under the 6th Framework Programme through the Key Action: Strengthening the European Research Area, Research Infrastructures, Contract No. RII3-CT-2003-505925. Work at Los Alamos was performed under the auspices of the U.S. DOE. Part of this work was carried out at the Brookhaven National Laboratory, which is operated for the U.S. Department of Energy by Brookhaven Science Associates (DE-Ac02-98CH10886).

\section*{References}

\bibliographystyle{unsrt}
\bibliography{Ce115_IOP}

\begin{thebibliography}{10}

\bibitem{Bia02}
A.~Bianchi, R.~Movshovich, N.~Oeschler, P.~Gegenwart, F.~Steglich, J.~D.
  Thompson, P.~G. Pagliuso, and J.~L. Sarrao.
\newblock \textrm{First-Order Superconducting Phase Transition in
  CeCoIn$_{5}$}.
\newblock {\em Phys. Rev. Lett.}, 89:137002, 2002.

\bibitem{Tay02}
T.~Tayama, A.~Harita, T.~Sakakibara, Y.~Haga, H.~Shishido, R.~Settai, and
  Y.~Onuki.
\newblock \textrm{Unconventional heavy-fermion superconductor CeCoIn$_{5}$: dc
  magnetization study at temperatures down to 50~mK}.
\newblock {\em Phys. Rev. B}, 65:180504, 2002.

\bibitem{Bia03a}
A.~Bianchi, R.~Movshovich, I.~Vekhter, P.~G. Pagliuso, and J.~L. Sarrao.
\newblock \textrm{Avoided Antiferromagnetic Order and Quantum Critical Point in
  CeCoIn$_{5}$}.
\newblock {\em Phys. Rev. Lett.}, 91:257001, 2003.

\bibitem{Pag03}
J.~Paglione, M.~A. Tanatar, D.~G. Hawthorn, E.~Boaknin, R.~W. Hill, F.~Ronning,
  M.~Sutherland, L.~Taillefer, C.~Petrovic, and P.~C. Canfield.
\newblock \textrm{Field-Induced Quantum Critical Point in CeCoIn$_{5}$}.
\newblock {\em Phys. Rev. Lett.}, 91:246405, 2003.

\bibitem{Pet01}
C.~Petrovic, P.~G. Pagliuso, M.~F. Hundley, R.~Movshovich, J.~L. Sarrao, J.~D.
  Thompson, Z.~Fisk, and P.~Monthoux.
\newblock \textrm{Heavy-fermion superconductivity in CeCoIn$_{5}$ at 2.3~K}.
\newblock {\em J. Phys.: Condens. Matter.}, 13:L337, 2001.

\bibitem{Iza01}
K.~Izawa, H.~Yamaguchi, Y.~Mastuda, H.~Shishido, Settai R., and Y.~Onuki.
\newblock \textrm{Angular Position of Nodes in the Superconducting Gap of
  Quasi-2D Heavy-Fermion Superconductor CeCoIn$_{5}$}.
\newblock {\em Phys. Rev. Lett.}, 87:057002, 2001.

\bibitem{Vor06}
A.~Vorontsov and I.~Vekhter.
\newblock \textrm{Nodal Structure of Quasi-Two-Dimensional Superconductors
  Probed by a Magnetic Field}.
\newblock {\em Phys. Rev. Lett.}, 96:237001, 2006.

\bibitem{Clo62}
A.~M. Clogston.
\newblock \textrm{Upper Limit for the Critical Field in Hard Superconductors}.
\newblock {\em Phys. Rev. Lett.}, 9:266, 1962.

\bibitem{Mak66}
K.~Maki.
\newblock \textrm{Effect of Pauli Paramagnetism on Magnetic Properties of
  High-Field Superconductors}.
\newblock {\em Phys. Rev.}, 148:362, 1966.

\bibitem{Orm02}
R.~J. Ormeno, A.~Sibley, C.~E. Gough, S.~Sebastian, and I.~R. Fisher.
\newblock \textrm{Microwave Conductivity and Penetration Depth in the Heavy
  Fermion Superconductor CeCoIn$_{5}$}.
\newblock {\em Phys. Rev. Lett.}, 88:047005, 2002.

\bibitem{Ful64}
P.~Fulde and R.~A. Ferrell.
\newblock \textrm{Superconductivity in a Strong Spin-Exchange Field}.
\newblock {\em Phys. Rev.}, 135:A550, 1964.

\bibitem{Lar64}
A.~I. Larkin and Y.~N. Ovchinnikov.
\newblock \textrm{Inhomogeneous State of Superconductors}.
\newblock {\em Sov. Phys. JETP}, 20:762, 1965.

\bibitem{Mat07}
Y.~Matsuda and H.~Shimahara.
\newblock \textrm{Fulde-Ferrell-Larkin-Ovchinnikov State in Heavy Fermion
  Superconductors}.
\newblock {\em J. Phys. Soc. Jpn.}, 76:051005, 2007.

\bibitem{Bia03b}
A.~Bianchi, R.~Movshovich, C.~Capan, P.~G. Pagliuso, and J.~L. Sarrao.
\newblock \textrm{Possible Fulde-Ferrell-Larkin-Ovchinnikov Superconducting
  State in CeCoIn$_{5}$}.
\newblock {\em Phys. Rev. Lett.}, 91:187004, 2003.

\bibitem{Rad03}
H.~A. Radovan, N.~A. Fortune, T.~P. Murphy, S.~T. Hannahs, E.~C. Palm, S.~W.
  Tozer, and D.~Hall.
\newblock \textrm{Magnetic Enhancement of Superconductivity from Electron Spin
  Domains}.
\newblock {\em Nature}, 425:51, 2003.

\bibitem{Kak05}
K.~Kakuyanagi, M.~Saitoh, K.~Kumagai, S.~Takashima, M.~Nohora, H.~Takagi, and
  Y.~Matsuda.
\newblock \textrm{Texture in the Superconducting Order Parameter of
  CeCoIn$_{5}$ Revealed by Nuclear Magnetic Resonance}.
\newblock {\em Phys. Rev. Lett.}, 94:047602, 2005.

\bibitem{Mit08}
V.~F. Mitrovi\'{c}, G.~Koutroulakis, M.~Klanj\ifmmode~\check{s}\else
  \v{s}\fi{}ek, M.~Horvati\ifmmode~\acute{c}\else \'{c}\fi{}, C.~Berthier,
  G.~Knebel, G.~Lapertot, and J.~Flouquet.
\newblock Comment on ``{T}exture in the {S}uperconducting {O}rder {P}arameter
  of {CeCoI}n$_{5}$ {R}evealed by {N}uclear {M}agnetic {R}esonance''.
\newblock {\em Phys. Rev. Lett.}, 101:039701, 2008.

\bibitem{Kum06}
K.~Kumagai, M.~Saitoh, T.~Oyaizu, Y.~Furukawa, S.~Takashima, M.~Nohara,
  H.~Takagi, and Y.~Matsuda.
\newblock \textrm{Fulde-Ferrell-Larkin-Ovchinnikov State in a Perpendicular
  Field of Quasi-Two-Dimensional CeCoIn$_{5}$}.
\newblock {\em Phys. Rev. Lett.}, 97:227002, 2006.

\bibitem{You07}
B.-L. Young, R.~R. Urbano, N.~J. Curro, J.~D. Thompson, J.~L. Sarrao, A.~B.
  Vorontsov, and M.~J. Graf.
\newblock {Microscopic Evidence for Field-Induced Magnetism in CeCoIn$_{5}$}.
\newblock {\em Phys. Rev. Lett.}, 98:036402, 2007.

\bibitem{Ken08}
M.~Kenzelmann, T.~Str\"{a}ssle, C.~Niedermayer, M.~Sigrist, P.~Padmanabhan,
  M.~Zolliker, A.~D. Bianchi, R.~Movshovich, E.~D. Bauer, J.~L. Sarrao, and
  J.~D. Thompson.
\newblock \textrm{Coupled Superconducting and Magnetic Order in CeCoIn$_{5}$}.
\newblock {\em Science}, 321:1652, 2008.

\bibitem{Esk03}
M.~R. Eskildsen, C.~D. Dewhurst, B.~W. Hoogenboom, C.~Petrovic, and P.~C.
  Canfield.
\newblock \textrm{Hexagonal and Square Flux Line Lattices in CeCoIn$_{5}$}.
\newblock {\em Phys. Rev. Lett.}, 90:187001, 2003.

\bibitem{DeB06}
L.~DeBeer-Schmitt, C.~D. Dewhurst, B.~W. Hoogenboom, C.~Petrovic, and M.~R.
  Eskildsen.
\newblock \textrm{Field Dependent Coherence Length in the Superclean,
  High-kappa Superconductor CeCoIn$_{5}$}.
\newblock {\em Phys. Rev. Lett.}, 97:127001, 2006.

\bibitem{Bia08}
A.~D. Bianchi, M.~Kenzelmann, L.~DeBeer-Schmitt, J.~S. White, E.~M. Forgan,
  J.~Mesot, M.~Zolliker, J.~Kohlbrecher, R.~Movshovich, E.~D. Bauer, J.~L.
  Sarrao, Z.~Fisk, C.~Petrovic, and M.~R. Eskildsen.
\newblock \textrm{Superconducting Vortices in CeCoIn$_{5}$: Toward the
  Pauli-Limiting Field}.
\newblock {\em Science}, 319:177, 2008.

\bibitem{Ohi08}
S.~Ohira-Kawamura, H.~Shishido, H.~Kawano-Furukawa, B.~Lake, A.~Wiedenmann,
  K.~Kiefer, T.~Shibauchi, and Y.~Matsuda.
\newblock \textrm{Anomalous Flux Line Lattice in CeCoIn$_{5}$}.
\newblock {\em J. Phys. Soc. Jpn.}, 77:023702, 2008.

\bibitem{Ich07b}
M.~Ichioka and K.~Machida.
\newblock \textrm{Vortex States in Superconductors with Strong
  Pauli-Paramagnetic Effect}.
\newblock {\em Phys. Rev. B}, 76:064502, 2007.

\bibitem{Chr77}
D.~K. Christen, F.~Tasset, S.~Spooner, and H.~A. Mook.
\newblock \textrm{Study of the Intermediate Mixed State of Niobium by
  Small-Angle Neutron Scattering}.
\newblock {\em Phys. Rev. B}, 15:4506, 1977.

\bibitem{Cle75}
J.~R. Clem.
\newblock \textrm{Simple Model for the Vortex Core in a Type-II
  Superconductor}.
\newblock {\em J. Low Temp. Phys.}, 18:427, 1975.

\bibitem{Yao97}
A.~Yaouanc, P.~Dalmas~de R\'{e}otier, and E.~H. Brandt.
\newblock \textrm{Effect of the Vortex Core on the Magnetic Field in Hard
  Superconductors}.
\newblock {\em Phys. Rev. B}, 55:11107, 1997.

\bibitem{Hao91}
Z.~Hao and J.~R. Clem.
\newblock \textrm{Reversible Magnetization and Torques in Anisotropic
  High-$\kappa$ Type-{II} Superconductors}.
\newblock {\em Phys. Rev. B}, 43:7622, 1991.

\bibitem{Bra95}
E.~H. Brandt.
\newblock \textrm{The flux-line lattice in superconductors}.
\newblock {\em Rep. Prog. Phys.}, 58:1465, 1995.

\bibitem{Yar97}
U.~Yaron, P.~L. Gammel, G.~S. Boebinger, G.~Aeppli, P.~Schiffer, E.~Bucher,
  D.~J. Bishop, C.~Broholm, and K.~Mortensen.
\newblock \textrm{Small Angle Neutron Scattering Studies of the Vortex Lattice
  in the UPt$_{3}$ Mixed State: Direct Structural Evidence for the
  $B\rightarrow{}C$ Transition}.
\newblock {\em Phys. Rev. Lett.}, 78:3185, 1997.

\bibitem{Kea00}
P.~G. Kealey, T.~M. Riseman, E.~M. Forgan, L.~M. Galvin, A.~P. Mackenzie, S.~L.
  Lee, D.~McK. Paul, R.~Cubitt, D.~F. Agterberg, R.~Heeb, Z.~Q. Mao, and
  Y.~Maeno.
\newblock \textrm{Reconstruction from Small-Angle Neutron Scattering
  Measurements of the Real Space Magnetic Field Distribution in the Mixed State
  of Sr$_{2}$RuO$_{4}$}.
\newblock {\em Phys. Rev. Lett.}, 84:6094, 2000.

\bibitem{For02}
E.~M. Forgan, S.~J. Levett, P.~G. Kealey, R.~Cubitt, C.~D. Dewhurst, and
  D.~Fort.
\newblock \textrm{Intrinsic Behavior of Flux Lines in Pure Niobium near the
  Upper Critical Field}.
\newblock {\em Phys. Rev. Lett.}, 88:167003, 2002.

\bibitem{Cub03}
R.~Cubitt, M.~R. Eskildsen, C.~D. Dewhurst, J.~Jun, S.~M. Kazakov, and
  J.~Karpinski.
\newblock \textrm{Effects of Two-Band Superconductivity on the Flux-Line
  Lattice in Magnesium Diboride}.
\newblock {\em Phys. Rev. Lett.}, 91:047002, 2003.

\bibitem{Gil04}
R.~Gilardi, J.~Mesot, S.~P. Brown, E.~M. Forgan, A.~Drew, S.~L. Lee, R.~Cubitt,
  C.~D. Dewhurst, T.~Uefuji, and K.~Yamada.
\newblock \textrm{Square Vortex Lattice at Anomalously Low Magnetic Fields in
  Electron-Doped Nd$_{1.85}$Ce$_{0.15}$CuO$_{4}$}.
\newblock {\em Phys. Rev. Lett.}, 93:217001, 2004.

\bibitem{Cub07}
R.~Cubitt, J.~S. White, M.~Laver, M.~R. Eskildsen, C.~D. Dewhurst, D.~McK.
  Paul, A.~J. Crichton, M.~Ellerby, C.~Howard, Z.~Kurban, and F.~Norris.
\newblock \textrm{Small-angle neutron scattering measurements of the vortex
  lattice in CaC$_{6}$}.
\newblock {\em Phys. Rev. B}, 75:140516, 2007.

\bibitem{Den09}
J.~M. Densmore, P.~Das, K.~Rovira, T.~D. Blasius, L.~DeBeer-Schmitt,
  N.~Jenkins, D.~McK.~Paul, C.~D. Dewhurst, S.~L. Bud'ko, P.~C. Canfield, and
  M.~R. Eskildsen.
\newblock \textrm{Small-Angle Neutron Scattering Study of the Vortex Lattice in
  Superconducting LuNi$_{2}$B$_{2}$C}.
\newblock {\em Phys. Rev. B}, 79:174522, 2009.

\bibitem{Ich99}
M.~Ichioka, A.~Hasegawa, and K.~Machida.
\newblock \textrm{Field dependence of the vortex structure in $d$-wave and
  $s$-wave superconductors}.
\newblock {\em Phys. Rev. B}, 59:8902, 1999.

\bibitem{Eil68}
G.~Eilenberger.
\newblock \textrm{Transformation of Gorkov's Equation for Type-II
  Superconductors into Transport-Like Equations}.
\newblock {\em Z. Phys.}, 214:195, 1968.

\bibitem{Not09}
$\mu \sim 0.67{H_{c2}^{Orb}}/{H_{c2}^{P}}$, where $H_{c2}^{P}$ is the
  Pauli-limited critical field observed at low temperatures, and $H_{c2}^{Orb}$
  is the expected orbital critical field extrapolated to $T=0$. $\mu$ is
  approximately half the Maki $\alpha$ parameter~\cite{StJ69} or 3.6 $\times$
  the parameter $\alpha_{para}$ used by Adachi~\emph{et al.}~\cite{Ada07}.

\bibitem{DeB07}
L.~DeBeer-Schmitt, M.~R. Eskildsen, M.~Ichioka, K.~Machida, N.~Jenkins, C.~D.
  Dewhurst, A.~B. Abrahamsen, S.~L. Bud'ko, and P.~C. Canfield.
\newblock \textrm{Pauli Paramagnetic Effects on Vortices in Superconducting
  TmNi$_{2}$B$_{2}$C}.
\newblock {\em Phys. Rev. Lett.}, 99:167001, 2007.

\bibitem{Set01}
R.~Settai, H.~Shishido, S.~Ikeda, Y.~Murukawa, M.~Nakashima, D.~Aoki, Y.~Haga,
  H.~Harima, and Y.~Onuki.
\newblock \textrm{Quasi-Two-Dimensional Fermi Surfaces and the de Haas-van
  Alphen Oscillation in Both the Normal and Superconducting Mixed States of
  CeCoIn$_{5}$}.
\newblock {\em J. Phys.: Condens. Matter}, 13:L627, 2001.

\bibitem{Ike01}
S.~Ikeda, H.~Shishido, M.~Nakashima, R.~Settai, D.~Aoki, Y.~Haga, H.~Harima,
  Y.~Aoki, T.~Namiki, H.~Sato, and Y.~Onuki.
\newblock \textrm{Unconventional Superconductivity in CeCoIn$_{5}$ Studied by
  the Specific Heat and Magnetization Measurements}.
\newblock {\em J. Phys. Soc. Jpn.}, 70:2248, 2001.

\bibitem{Ich09}
M.~Ichioka and K.~Machida.
\newblock {\em \emph{(Private Communication)}}, 2009.

\bibitem{Sey08}
G.~Seyfarth, J.~P. Brison, G.~Knebel, D.~Aoki, G.~Lapertot, and J.~Flouquet.
\newblock \textrm{Multigap Superconductivity in the Heavy-Fermion System
  CeCoIn$_5$}.
\newblock {\em Phys. Rev. Lett.}, 101:046401, 2008.

\bibitem{Ozc03}
S.~\"{O}zcan, D.~M. Broun, B.~Morgan, R.~K.~W. Haselwimmer, J.~L. Sarrao,
  S.~Kamal, C.~P. Bidinosti, P.~J. Turner, M.~Raudsepp, and J.~R. Waldram.
\newblock \textrm{London penetration depth measurements of the heavy-fermion
  superconductor CeCoIn$_{5}$ near a magnetic quantum critical point}.
\newblock {\em Europhys. Lett.}, 62:412, 2003.

\bibitem{Gru66}
L.~W. Gruenberg and L.~Gunther.
\newblock \textrm{Fulde-Ferrell Effect in Type-II Superconductors}.
\newblock {\em Phys. Rev. Lett.}, 16:996, 1966.

\bibitem{Ada03}
H.~Adachi and R.~Ikeda.
\newblock \textrm{Effects of Pauli Paramagnetism on the Superconducting Vortex
  Phase Diagram in Strong Fields}.
\newblock {\em Phys. Rev. B}, 68:184510, 2003.

\bibitem{Esk98}
M.~R. Eskildsen, K.~Harada, P.~L. Gammel, A.~B. Abrahamsen, N.~H. Andersen,
  G.~Ernst, A.~P. Ramirez, D.~J. Bishop, K.~Mortensen, D.~G. Naugle, K.~D.~D.
  Rathnayaka, and P.~C. Canfield.
\newblock \textrm{Intertwined Symmetry of the Magnetic Modulation and the
  Flux-Line Lattice in the Superconducting State of TmNi$_{2}$B$_{2}$C}.
\newblock {\em Nature}, 393:242, 1998.

\bibitem{Hia08}
N.~Hiasa and R.~Ikeda.
\newblock \textrm{Instability of Square Vortex Lattice in $d$-Wave
  Superconductors is due to Paramagnetic Depairing}.
\newblock {\em Phys. Rev. Lett.}, 101:027001, 2008.

\bibitem{StJ69}
D.~Saint-James~\textit{et al.}
\newblock {\em Type-II Superconductivity}.
\newblock Pergamon, New York, 1969.

\bibitem{Ada07}
H.~Adachi, M.~Ichioka, and K.~Machida.
\newblock \textrm{Mixed-State Thermodynamics of Superconductors with Moderately
  Large Paramagnetic Effects}.
\newblock {\em J. Phys. Soc. Jpn.}, 74:064502, 2007.

\end{thebibliography}
\newpage
\begin{figure}[t]
\centering
  \includegraphics[width=12cm]{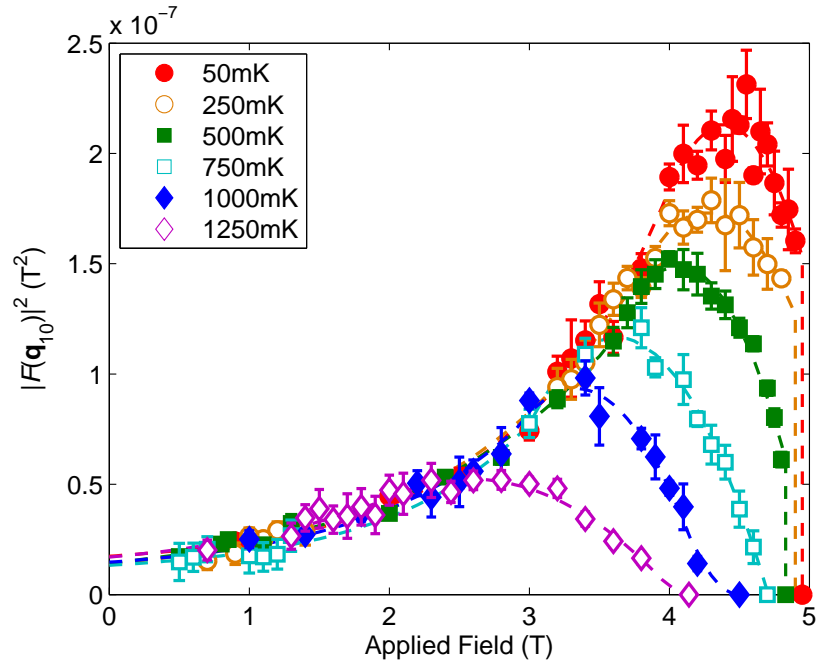}
  \caption{Field-dependence of the square of the flux line lattice form factor for first-order Bragg reflections, $|F(\textbf{q}_{10})|^{2}$, for temperatures up to 1250~mK. The dashed lines are guides-to-the-eye. The points at which the form factor goes to zero are obtained from measurements of $H_{c2}$~\cite{Bia03b}.}
\end{figure}

\newpage
\begin{figure}
\centering
  \includegraphics[width=12cm]{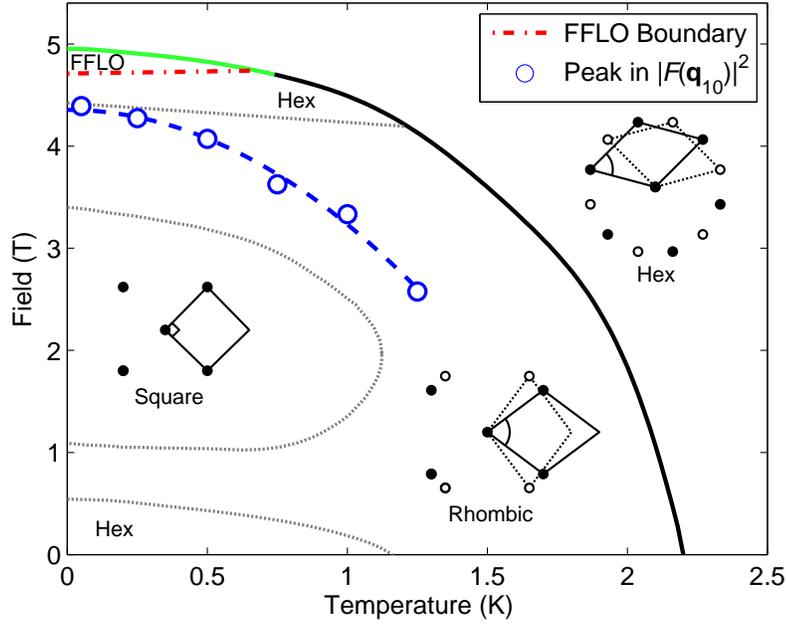}
  \caption{Superconducting phase diagram for CeCoIn$_{5}$ with $H \parallel$\textbf{c}, indicating the regions occupied by the different flux line lattice (FLL) structures reported in \cite{Esk03,DeB06,Bia08,Ohi08}. Phase boundaries separating regions of different FLL structure are shown by gray dotted lines. The inset diagrams show the three FLL configurations. Here the crystalline [100] direction is vertical, and first-order Bragg spots belonging to different FLL domains are denoted by open and filled circles respectively. The green and black solid lines show where the superconducting to normal state transition is first- and second-order respectively. The open blue circles show the temperature evolution of the maximum of the form factor ($|F(\textrm{\textbf{q}}_{10})|^2$) , with the blue dashed line being a guide-to-the-eye. The error bars of the data points are comparable in size to the data symbol. The area enclosed by the green solid and red dash-dot lines provide an estimate of the region occupied by the FFLO phase, as deduced from data shown in \cite{Bia03b,Kum06}.}
\end{figure}

\newpage
\begin{figure}
\centering
  \includegraphics[width=12cm]{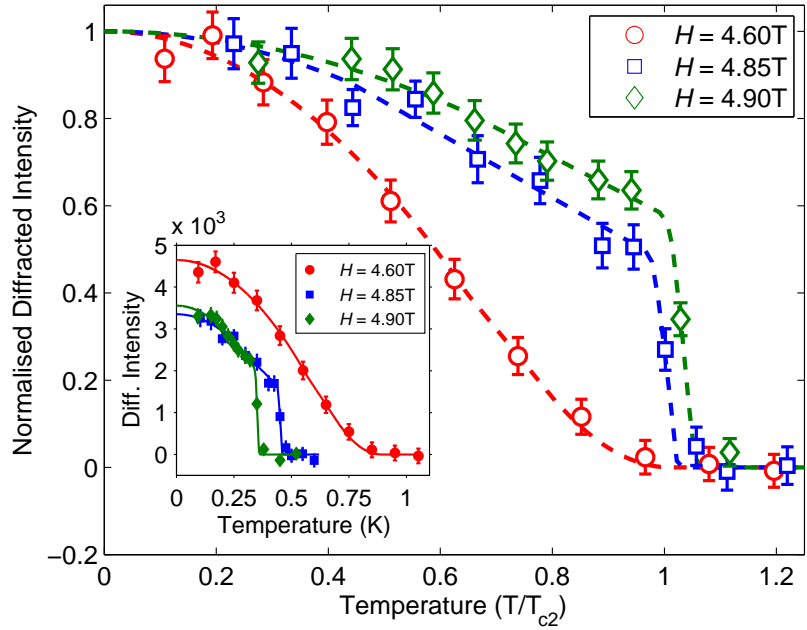}
  \caption{The normalised temperature-dependence of the diffracted intensity at the peak of the rocking curve for a first-order hexagonal flux line lattice Bragg spot lying along $\langle110\rangle$. Data were collected at fields of 4.60~T, 4.85~T and 4.90~T. The dashed lines are guides to the eye. The inset shows the same data as the main figure, but in absolute (un-normalised) units of counts per standard monitor. Since the rocking curve width is resolution limited, these are $\propto~I_{10}$ in Eqn. (1).}
\end{figure}

\newpage
\begin{figure}
\centering
  \includegraphics[width=12cm]{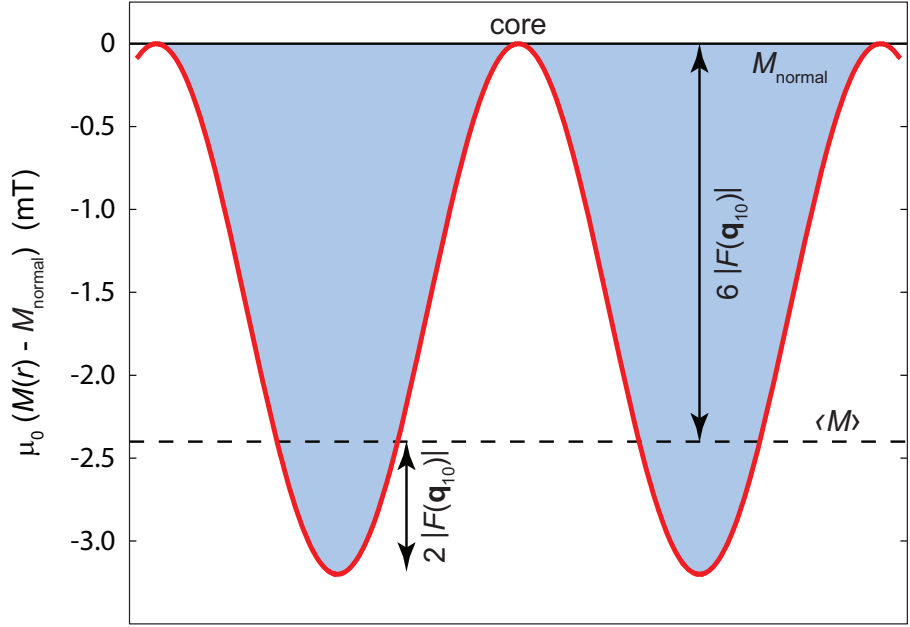}
  \caption{A schematic of the spatial variation of the local magnetisation along the nearest neighbor direction between adjacent flux line cores, and for an applied field ($H_{\textrm{appl.}}$) just below $H_{c2}$. The magnetisation is expected to be close to the the normal state value in the cores, and decreases between the flux lines where the Pauli paramagnetism is suppressed by the Cooper pairing of the supercarriers. The local magnetisation varies from $+6 |F(\textbf{q}_{10})|$ to $-2 |F(\textbf{q}_{10})|$ relative to $\mu_0 \langle M \rangle$. The minimum in the local magnetisation is $\mu_0 \langle M \rangle - 3 |F(\textbf{q}_{10})|$ and is obtained along the FLL next nearest neightbor direction (unit cell diagonal) (not shown). The shading indicates the origin of the reduced magnetisation in the superconducting state.}
\end{figure}

\end{document}